%% file: paper.tex
\documentclass{ppig}
\usepackage{epsfig}
\usepackage{booktabs}
\usepackage{ucs}
\usepackage[utf8x]{inputenc}
\usepackage{apacite}

\usepackage{xcolor}
\usepackage{listings}

\usepackage{soul}
\usepackage{xspace}
\usepackage{url}
\usepackage{graphicx}
\usepackage{listings,multicol}
\usepackage[caption=false]{subfig}  
\usepackage[export]{adjustbox}
\usepackage{amsmath}
\usepackage{textcomp}
\usepackage{breakurl} 
\usepackage{cleveref}
\usepackage{algorithm}
\usepackage{xparse}
\usepackage[noend]{algpseudocode}
\usepackage{balance}
\usepackage{roboto}
\renewcommand{\texttt}[1]{\textcolor{blue}{\fontfamily{Roboto}\selectfont #1}}

\newcounter{observation}
\newcommand{\observation}[1]{\refstepcounter{observation}
	\begin{center}
		\framebox{
			\begin{minipage}{0.95\textwidth}
				{} \textit{#1}
			\end{minipage}
		}
	\end{center}
}

\newcommand{\eins}{{Einstellung}\xspace}

\definecolor{APA_stats}{RGB}{100, 100, 120}
\newcommand{\APAstats}[2]{\textcolor{APA_stats}{(M = #1, SD = #2)}}
\newcommand{\APAstatsNoBraces}[2]{\textcolor{APA_stats}{M = #1, SD = #2}}

\newcommand{\APAt}[3]{\textcolor{APA_stats}{T(#1) = #2, p = #3}}
\newcommand{\APAh}[3]{\textcolor{APA_stats}{H(#1) = #2, p = #3}}
\newcommand{\APAr}[3]{\textcolor{APA_stats}{r(#1) = #2, p = #3}}

\title{Overcoming the Mental Set Effect in Programming Problem Solving}

\author{Agnia Sergeyuk, Sergey Titov, Yaroslav Golubev, Timofey Bryksin \\
JetBrains Research \\
\{agnia.sergeyuk, sergey.titov, yaroslav.golubev, timofey.bryksin\}@jetbrains.com}

\date{30.05.2023}

\begin{document}
\maketitle
\thispagestyle{empty}

\begin{abstract}
This paper adopts a cognitive psychology perspective to investigate the recurring mistakes in code resulting from the mental set (Einstellung) effect. The Einstellung effect is the tendency to approach problem-solving with a preconceived mindset, often overlooking better solutions that may be available. This effect can significantly impact creative thinking, as the development of patterns of thought can hinder the emergence of novel and creative ideas. Our study aims to test the Einstellung effect and the two mechanisms of its overcoming in the field of programming. The first intervention was the change of the color scheme of the code editor to the less habitual one. The second intervention was a combination of instruction to ``forget the previous solutions and tasks'' and the change in the color scheme. During the experiment, participants were given two sets of four programming tasks. Each task had two possible solutions: one using suboptimal code dictated by the mental set, and the other using a less familiar but more efficient and recommended methodology. Between the sets, participants either received no treatment or one of two interventions aimed at helping them overcome the mental set. The results of our experiment suggest that the tested techniques were insufficient to support overcoming the mental set, which we attribute to the specificity of the programming domain. The study contributes to the existing literature by providing insights into creativity support during problem-solving in software development and offering a framework for experimental research in this field.
\end{abstract}

\input{sections/01-intro}
\input{sections/02-background}
\input{sections/03-method}
\input{sections/04-findings}
\input{sections/05-discussion}
\input{sections/06-ttv}
\input{sections/07-conclusion}

\balance

\bibliography{paper}
\bibliographystyle{apacite} 
\end{document}

%% file: sections/01-intro.tex
\section {INTRODUCTION} \label{sec:introduction}

In recent years, there has been a growing trend of applying knowledge from cognitive psychology to address problems in software engineering~\shortcite{lenberg2015behavioral,bidlake2020systematic,mangalaraj2014distributed}. This approach has shown its effectiveness in various areas, such as fault detection in software inspections~\shortcite{anu2016using}, understanding the role of the cognitive load in program comprehension and maintenance~\shortcite{fakhoury2018effect}, and exploring the neural correlates of programming tasks~\shortcite{floyd2017decoding}, among others. In this paper, we adopt a cognitive psychology perspective to examine habitual mistakes in code, aiming to shed light on the role of cognitive processes in software development tasks. By leveraging insights from cognitive psychology, we seek to deepen our understanding of the underlying cognitive mechanisms that contribute to coding errors and explore potential interventions to improve coding performance and foster creativity in software development tasks.

Habitual mistakes, \textit{i.e.,} mistakes driven by a fixation on previous knowledge, in the context of programming can result in maintenance issues and software bugs~\shortcite{glass2003facts}. Some of these recurring mistakes or error-prone points in code may be attributed to the mental set (Einstellung) effect --- a cognitive rigidness that occurs during problem-solving when a habitual solution is applied without considering other potentially more optimal and suitable alternatives, particularly in familiar contexts~\shortcite{luchins1942mechanization}. The mental set effect can significantly impact creative thinking, as the development of patterns of thought can hinder the emergence of novel ideas that may be better suited for the task at hand~\shortcite{davis1999barriers,wiley1998expertise,kilgour2006improving}. 

In the existing research, techniques for overcoming the \eins effect were studied in a laboratory on relatively abstract mathematical tasks in the form of water-jar problems, or anagrams, labyrinths, etc.~\shortcite{chen2004schema,lovett1996history,lee2004creative}. It was shown that one of the possible ways to weaken the effect is a change in the direction of awareness and cognitive control: its enhancement on the target task or refocusing on an additional task~\shortcite{tuchtieva2016,tukhtieva2014influence}. Research also show that people who were instructed to intentionally forget the habituated solution of some problem relied less on it, \textit{i.e.}, the \eins effect weakened for them~\shortcite{tempel2019directed,storm2010overcoming}. Our aim was to test the \eins effect and the mechanisms of its overcoming in more real-world circumstances and in the field of programming.

Our study investigated the \eins effect in a sample of 129 Python developers. We conducted an online pre-experimental survey to identify participants who exhibited habitual mistakes (mental sets) that were relevant to our research. Among them, 57 individuals (with a mean experience in Python of 25 months) demonstrated at least one mental set. The final sample for our study consisted of 39 participants who successfully completed all experimental procedures. 

For the experiment, we developed an online Flask\footnote{Flask framework: \url{https://flask.palletsprojects.com/en/2.3.x/}} application with an embedded Ace9 code editor\footnote{Ace9 code editor: \url{https://ace.c9.io/}} serving as the programming environment. In this environment, participants were presented with two sets of tasks and one of three interventions between those sets. The tasks were carefully selected to specifically test the \eins effect, with each task having at least two solutions: one habitual (indirect) solution, and one "correct" (direct) solution that is recommended in style guides of the used programming language and more efficient in terms of code execution time. Participants were assigned to one of three groups: \textbf{Control} (no intervention), \textbf{Change} (task-irrelevant change in the color scheme of the environment), or \textbf{Change and Forget} (the same change in the color scheme plus an additional "forget cue" instruction). Following the experiment, participants were asked to rate the helpfulness and comfortability of the intervention they received (if any) on a scale from 0 to 5, and provide reasons for switching or not switching from the indirect solution to the direct one during the second set of tasks. 

The results of our study shed light on the effectiveness of the tested interventions in overcoming mental sets among Python developers. Despite the participants' reported engagement in the process of rethinking their solutions (in 54\% of cases), our findings indicate that the provided interventions were unhelpful (\APAstatsNoBraces{0.97}{1.18} on a scale from 0 to 5) and insufficient for the switch from habitual indirect solutions to the recommended direct solutions.

In summary, our study contributes to the existing literature by providing insights into creativity support during problem-solving in software development, offering a framework for experimental research, and sharing open-source code and data for future investigations in this field.\footnote{All the code used for the experiment and the data analysis is publicly available on Zenodo: \url{https://doi.org/10.5281/zenodo.5893501}}

%% file: sections/02-background.tex
\section {BACKGROUND} \label{sec:background}

\medskip
\subsection{The \eins Effect and Creativity}

The \eins effect, also known as the mental set effect, was originally demonstrated by Luchins in 1942~\shortcite{luchins1942mechanization}. In Luchins' experiments, participants were given a series of 11 problems that involved imagining using three water jars of different volumes to measure a certain amount of water. The tasks were divided into two sets. The first set had a fixed solution --- \textit{e.g.}, getting 100 liters with 21, 127, and 3 liters jars might be solved as $(127 - 21 - 2*3)$, while the second set could be solved using either the same approach or a more direct and simpler approach --- \textit{e.g.}, the task to get 18 liters with 15, 39, and 3 liters jars might be solved in two ways: a habituated one $(39 - 15 - 2*3)$ or a simple one $(15 + 3)$. The habituated approach, referred to as the \textbf{indirect} solution, required more time and effort, while the direct approach --- \textbf{direct} solution --- was quicker and less habitual but still known. The findings showed that direct approaches to solving the second set of tasks were less frequent compared to indirect approaches, indicating the presence of the mental set effect, which was operationalized as a difference in frequencies of indirect and direct solutions~\shortcite{guetzkow1951analysis}.

Subsequent studies replicated Luchins' findings~\shortcite{mckelvie1990einstellung} and applied the concept of the mental set to different types of problems: some of them did not require specific prior knowledge or skill~\shortcite{chen2004schema,lovett1996history,lee2004creative}, and in others, the mental set was assumed to be already formed based on the knowledge stored in the long-term memory~\shortcite{bilalic2008inflexibility,wiley1998expertise,saariluoma1992error}. 

The mental set was proven to inhibit creativity by leading individuals to approach new problems with preconceived notions or rigid thinking patterns based on past experiences or familiar solutions~\shortcite{wiley1998expertise,kilgour2006improving}. Therefore, overcoming the mental set may be necessary to unlock creative and innovative solutions.

In our current research, we investigate the \eins effect from the perspective of the Adaptive Control of Thought theory (ACT-R), which posits that the effect is a result of selection mechanisms in human cognition~\shortcite{anderson1993production,ritter2019act}. ACT-R is a comprehensive theory of cognition that seeks to explain how humans acquire, represent, and process knowledge to perform various cognitive tasks, such as problem solving. According to this framework, the likelihood of selecting a solution for a given problem depends on the history-of-success and distance-to-goal rates that are internally attributed to the solution, and this likelihood changes as these rates change~\shortcite{lovett1996history,ollinger2008investigating}. Therefore, overcoming the mental set may be achieved by proving the solution unsuccessful in terms of the task or modifying the characteristics of the task's goal. Previous studies on overcoming the \eins effect have mainly been conducted in laboratory settings using abstract mathematical tasks, such as water-jar problems or anagrams. However, our goal is to investigate the effect and its mechanisms in more realistic contexts, specifically in the field of programming.

In the context of software engineering, the \eins effect can be demonstrated in coding practices and code style. Unlike bugs or logical errors, coding style issues may not have an immediate effect on the functionality of the code, which can lead to them becoming ingrained habits among developers. As a result, developers may struggle with adopting suggested code style guidelines, even when presented with feedback or recommendations. This can lead to error-prone code and reduced maintainability~\shortcite{martin2009clean,glass2003facts}. To address this issue, several tools have been developed to detect and highlight coding style problems, and in some cases, even automatically fix them~\shortcite{smirnov2021revizor,wiese2017teaching,birillo2022hyperstyle,blau2015frenchpress}. These tools aim to provide direct feedback to developers to help them identify and correct coding style issues. However, these tools often do not take into account known cognitive mechanisms that underlie the mental set effect.

\medskip
\subsection{Overcoming of the \eins Effect} \label{overcoming}

Our research aimed to test two approaches that have been proposed in prior works to overcome the mental set effect~\shortcite{tuchtieva2011, tukhtieva2014influence, vallee2011einstellung, tempel2019directed, storm2010overcoming}. According to previous research, the \eins effect may arise due to the mechanization of a particular solution, where the problem solver becomes automatic in their approach. To overcome this effect, one needs to shift from automaticity to consciously controlled actions. This can be achieved by triggering a reevaluation of the solution's history-of-success and distance-to-goal rates~\shortcite{lovett1996history}, and by restructuring the situation in terms of connections and relationships between objects~\shortcite{wertheimer1959productive}. 

One way to de-automatize cognitive processes is to \textbf{diversify task conditions}, which can affect distance-to-goal rates~\shortcite{lovett1996history}, stimulate conscious activity, and facilitate generation and testing of different hypotheses~\shortcite{allakhverdov2009reflection}. For instance, systematic changes in the irrelevant parameters of the task can complicate it, leading to the activation of conscious control~\shortcite{allakhverdov2008awareness}. This implies that changing aspects of the task that are irrelevant to the solution could be a potential strategy to overcome mechanization or the mental set effect~\shortcite{tukhtieva2014influence,vallee2011einstellung,luchins1950new}. In the context of the water-jar problem, the given volume of water is a relevant aspect, but the way the jars are presented, such as their color or whether they are physical or virtual, does not affect the solution~\shortcite{luchins1950new}. Tukhtieva~\shortcite{tuchtieva2011,tukhtieva2014influence,tuchtieva2016} conducted detailed studies on the effect of task-irrelevant changes on overcoming the mental set and found that when water jars were presented as digital images with a dramatic change in the background of the task, such as from a neutral one-colored background to bright and colorful photographs, it reduced the effect of the mental set and helped individuals to overcome it.

Another approach to trigger reevaluation of solutions and overcome the mental set is through \textbf{intentional forgetting}, which involves giving direct instruction to "forget the tasks and solutions from before"~\shortcite{tempel2019directed,storm2010overcoming} and which might affect the internal history-of-success rating~\shortcite{lovett1996history}. This technique is adapted from research on intentional forgetting using the list method~\shortcite{basden1993directed}. In the list method, two lists of words, \textit{e.g.}, furniture pieces, are sequentially presented. The first list is presented with an instruction to remember it, and after the participant memorized it, the list is marked with an instruction to forget it under some pretext, such as defining the list as needed only for warm-up. Next, the second list is presented with an instruction to remember it. After a certain period of time, the subject's ability to recall the material from both lists is tested. This method is designed to label information that should not be processed and, therefore, remembered, requiring the suppression of such information using cognitive control. This process, as suggested by previously mentioned research, may be effective in overcoming the mental set effect.

We took a unique approach by combining the existing experimental paradigm of the \eins effect testing with the known intervention techniques and applied them in the field of programming. This novel application allowed us to explore uncharted territory and draw conclusions about the domain specificity and toolification potential of cognitive phenomena. By bridging the gap between cognitive psychology and programming, we aimed to uncover new insights and contribute to the understanding of how cognitive processes influence creative problem-solving in programming tasks.

%% file: sections/03-method.tex
\section{METHODOLOGY} \label{sec:method}

In our study, we utilized Luchins' paradigm~\shortcite{luchins1942mechanization} and adapted it for coding tasks in order to compare two approaches for overcoming the \eins effect in a sample of developers. The first approach involved \textbf{diversifying task conditions} by changing a task-irrelevant aspect, specifically the color scheme of the development environment. The second approach included an \textbf{intentional forgetting} cue along with the color scheme change. 

To investigate the effectiveness of the interventions in overcoming the \eins effect, we formulated the following research questions:
\begin{itemize}
    \item \textbf{RQ1.} What is the prevalence of the \eins effect among Python developers?
    \item \textbf{RQ2.} Do the provided interventions facilitate a shift in problem-solving strategies from indirect to direct, indicating the overcoming of the mental set?
    \item \textbf{RQ3.} Do the provided interventions facilitate solution de-automatization?
    \item \textbf{RQ4.} Do the reported helpfulness and comfortability scores, as measures of the usability of the interventions, correlate with each other and with the overcoming of the mental set?
\end{itemize}

In this section, we provide a detailed description of the sample, materials, interventions, data collection, and analysis procedures. First, we outline the sample characteristics. Next, we describe the study materials, which include the adapted programming tasks serving as a problem-solving proxy, as well as the interventions applied to overcome the \eins effect. We then detail the data collection process and the analysis procedures employed to address the formulated research questions.

\medskip
\subsection{Sample}

\begin{figure}[t]
    \includegraphics[width=\columnwidth]{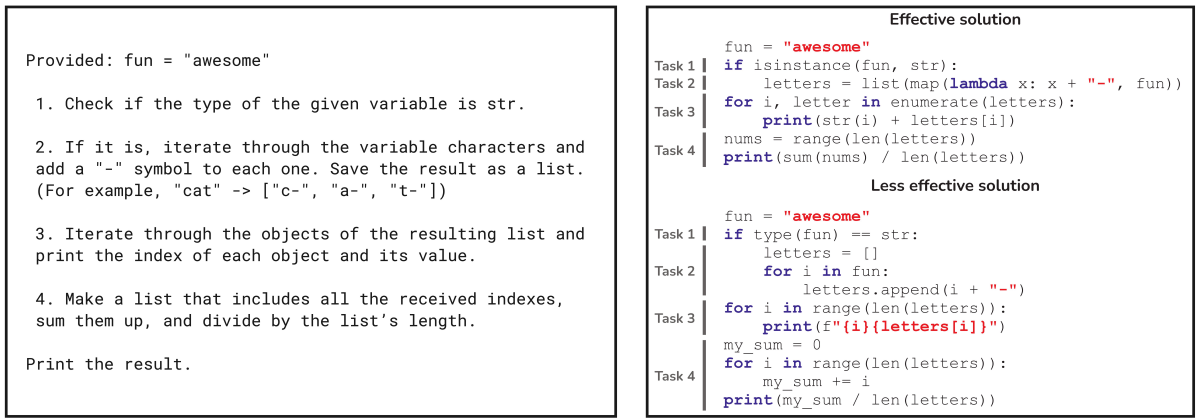}
    \caption{The screening task and its solutions.}
    \label{fig:two-tasks}
\end{figure}

To gather our sample, we administered a screening questionnaire to identify Python developers who consistently use less efficient (indirect) code, despite being aware of more direct and efficient alternatives, indicating a mental set for tasks relevant to our research. The questionnaire included informed consent and inquiries about participants' experience with Python in months. Participants then completed a programming task (see Figure~\ref{fig:two-tasks}). This task consisted of four steps, which were counterparts to tasks used in the subsequent experiment. They were also asked to select familiar Python functions from a provided list, including both target functions (considered indirect and direct in our research) and other functions not used in the study serving as noise. This allowed us to assess whether participants used the indirect function in the tasks, despite being familiar with the direct function based on the checklist.

We extended invitations to participate in the primary stages of the study to respondents who exhibited a mental set for the target Python functions. Specifically, we invited developers who satisfied both of the following criteria: (a) their responses to the given problem contained indirect solutions, as presented in Figure~\ref{fig:two-tasks}, and (b) they indicated familiarity with both indirect and direct functions in the checklist. This yielded an invitation to 57 out of 129 respondents who completed the questionnaire to participate in the primary experimental part of the research. However, not all the invitees could complete this step, resulting in a final sample size of 39 individuals.

\medskip
\subsection{Materials}\label{subsec:materials}

The crucial part of our experiment was finding the appropriate programming problems that could serve as the \eins tasks. It was decided to focus on the mistakes in code style rather than bugs, as the former don't make the code un-executable and due to that might be easily habituated. We identified four Python problems from the \textit{Wemake-python-styleguide}\footnote{Wemake-python-styleguide: \url{https://wemake-python-stylegui.de/en/latest/}} that could be solved in two different ways, one of which was more efficient time-wise and recommended as a ``correct'' one in style guides: 
\begin{itemize}
    \item \textbf{Iterate through the list and return every item's index}. This task might be solved using either \texttt{enumerate} or \texttt{range}, the first option is preferred.
    \item \textbf{Compare types}. Two possible solutions for this task imply using the \texttt{isinstance} or \texttt{type} functions, the first option is preferred.
    \item \textbf{Sum things}. \textit{Wemake-python-styleguide} indicates that for such tasks, using \texttt{for} loops with the \texttt{+=} assigning operator inside indicates that one iteratively sums elements of the collection. However, this is what the built-in \texttt{sum} function does. Thus, the latter should be used. 
    \item \textbf{Iterate through a string}. One could do this (1) with a loop, (2) using list comprehension, or (3) using the built-in \texttt{lambda} function. The latter is more straightforward. However, it is recommended to use more readable list comprehension. That is why we marked only the loop-based solution as an indirect one, and the other two were considered direct.
\end{itemize}
A more detailed discussion of the recommended variants and a comparison of their performance can be found in our online appendix available on Zenodo.\footnotemark[1]

Further in the text, we use the names of direct (in terms of our research) functions as labels of the corresponding mental set types (\textit{e.g.}, \texttt{enumerate} as a label of the mental set for \texttt{range}).

To effectively assess the impact of experimental treatments on mental set overcoming, we designed two sets of four previously described programming tasks. Each task in both sets required participants to provide solutions that could be attained through either direct or indirect functions. The second set of tasks was intentionally structured to have the same underlying structure and require the same computational logic as the tasks in the first set, but the wording and input data were changed. This approach allowed us to make an apples-to-apples comparison of the solutions provided by participants across various treatments and reach definitive conclusions about their ability to overcome the mental set.

As a medium for our experiment, we developed an online Flask application using the Python language with an embedded Ace9 code editor as the programming environment.

\medskip
\subsection{Intervention} \label{intervention}

To ensure ecological validity~\shortcite{orne1968ecological}, our experimental treatment aimed to replicate real-life situations. To achieve this, we selected task parameters and cues for the intervention that could be easily implemented in an Integrated Development Environment (IDE) without significant modifications, either through a simple plugin or existing environmental features. Previous research has used irrelevant changes, such as background color or representational form changes in various tasks, from simple math equations to anagram solving and memory tasks~\shortcite{moroshkina2008soznatelnyy,gershkovich2011explicit,tukhtieva2014influence,vallee2011einstellung,luchins1950new}. Drawing insights from this research, we adapted and applied similar principles in our study. As a result, we formed three intervention groups:
\begin{itemize}
    \item The \textbf{Change} group was presented with only the task-irrelevant change. Specifically, we changed the color scheme of the coding environment from the habitual Light or Dark to the contrast one, using the two most common, standard, and most contrast themes of the embedded Ace9 code editor: \textit{Eclipse} as the light one and \textit{Monokai} as the dark one. This change was selected as it is available in most coding environments and is not dependent on the task or the editor, making it easily implementable. Additionally, such a change is similar to a real-life situation where people switch their gadget's theme from light to dark and vice versa.
    \item The \textbf{Change and Forget} group received the same change as the Change group, along with the forget cue. The forget cue was designed based on the classical list-method paradigm used in forgetting research and was adapted to our study's goals. Specifically, participants in this group were instructed to forget their previous solutions and tasks as they were part of the warm-up phase, and to proceed to the main stage.
    \item The \textbf{Control} group received no intervention except the instruction to continue.
\end{itemize}

\medskip
\subsection{Data Collection}

\begin{figure*}
    \centering
    \includegraphics[width=\textwidth]{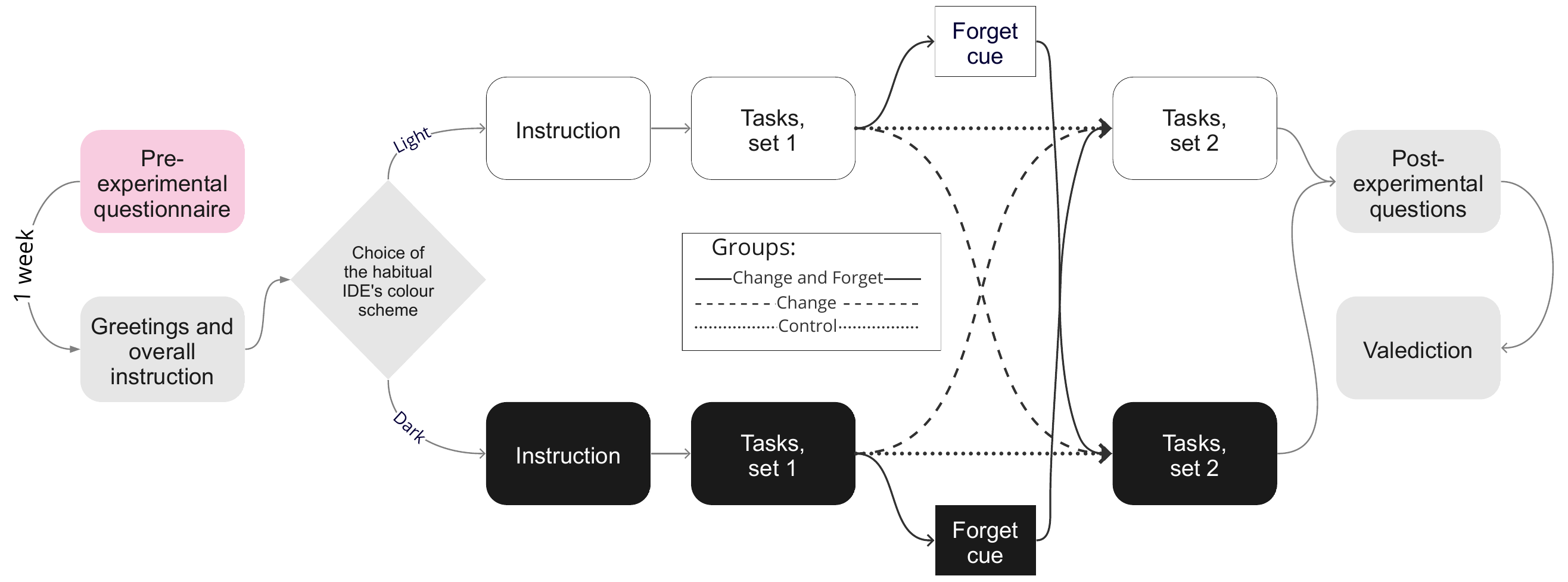}
    \caption{The procedure of the experiment.}
    \label{fig:pipeline}
\end{figure*}

The experimental procedure, depicted in Figure~\ref{fig:pipeline}, is described in detail below:
\begin{itemize}
    \item After a delay of one week from the day the screening questionnaire was filled out (to allow the memory trace of the questionnaire and its tasks to cool down), we sent an email to the selected participants containing a URL to access the experiment asynchronously.
    \item As the first step, participants were asked to choose their habitual color scheme (Light or Dark) for their programming environment. This was done to compare solutions created in an environment where the mental set had been repeatedly habituated and used, with solutions generated under relatively new circumstances.
    \item Next, participants were presented with a general experiment description stating, "In the next 10 minutes, you will have to solve a set of problems in Python. We will not check the correctness of the code, but it is important to us that the solution to the problem is as algorithmically efficient as possible. Please use the most efficient code to solve the problem." 
    \item Following this, a randomized sequence of four programming problems, as described in Section~\ref{subsec:materials}, was presented in the chosen color scheme. 
    \item Following this step, the participants received an intervention according to the group to which they were assigned: Change, Change and Forget, or Control, as described earlier in the section. 
    \item After that, the second set of programming problems was provided. The set consisted of four problems similar to those in the first set, but with the altered wording containing different input data to manipulate.
    \item Finally, an open-ended question was presented to determine the reasons for applying a particular type of solution for the second set of tasks. For the groups with the change of the task-irrelevant aspect, two 5-point Likert scales~\shortcite{likert1934simple} were presented to mark the comfortability level of the change in the presented color scheme (if any), as well as the level of its helpfulness in terms of solving the second set of tasks.
\end{itemize}

During the experiment, we gathered various data such as the participants' preferred habitual color scheme, the solutions they generated for each task, and their response time for each task. A response time difference served as a proxy measure of the cognitive load, with a larger difference indicating higher cognitive processing demands~\shortcite{rheem2018use}. Additionally, we assigned each participant a label indicating their group and the corresponding mental set effect they exhibited.

\medskip
\subsection{Data Analysis}

All the obtained data was analyzed in Python. Statistical analysis was performed using Scipy~\shortcite{virtanen2020scipy}, NumPy~\shortcite{harris2020array}, Pandas~\shortcite{mckinney2010data}, and Pingouin~\shortcite{vallat2018pingouin}. 

To correspond with the research questions, the following statistical hypotheses were formulated:
\begin{itemize}
    \item \textbf{H1.} There is a statistically significant difference in the experience between individuals who exhibited the \eins effect and those who did not. An independent T-Test was conducted on the initial questionnaire data to test this hypothesis.
    \item \textbf{H2.} The provided interventions result in a significant shift in problem-solving strategies from indirect to direct, indicating the overcoming of the mental set. Data visualization was performed to examine if there is any solution change.
    \item \textbf{H3.} The provided interventions lead to a significant difference in response time between corresponding blocks of tasks. The Kruskal-Wallis H test was utilized to determine whether there is an effect of group on the response time difference.
    \item \textbf{H4.} The reported helpfulness and comfortability scores are significantly correlated with each other and with the overcoming of the mental set. To examine these correlations, we used Spearman's correlation coefficient.
\end{itemize}

%% file: sections/04-findings.tex
\section{RESULTS} \label{sec:findings}

\medskip
\subsection{Prevalence of the Mental Set (RQ1, H1)} 

\begin{figure}[h]
    \centering
    \includegraphics[width=0.5\textwidth]{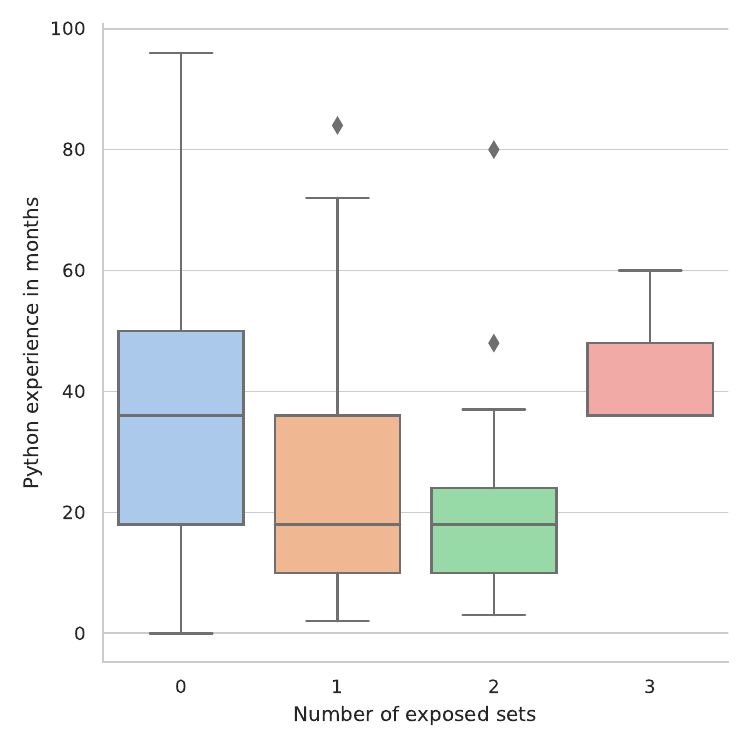}
    \caption{Number of sets exposed by participants of different months of experience.}
    \label{fig:months}
\end{figure}

Our analysis showed that approximately 44\% of the participants in our sample exhibited the mental set. Among the 129 individuals who took part in the screening questionnaire, 57 showed evidence of using less efficient, indirect functions to solve programming tasks despite being aware of more efficient, direct functions. Notably, we found a statistically significant difference in the number of months of Python experience between individuals who exhibited the mental set and those who did not, with those displaying the mental set having an average of 25 months of experience compared to an average of 37 months for those without the mental set (\APAt{126.998}{2.806}{.006}). This finding suggests that programmers with less experience are more prone to producing code with errors. This increased error rate can be attributed to their tendency to rely heavily on habituated knowledge, resulting in a fixation on familiar solutions without considering alternative approaches.

In terms of the types of mental set displayed, 41 participants exhibited only one type, 13 exhibited two types, and 3 exhibited three types, with none displaying all four types (see \Cref{fig:months}). The mental set related to the \texttt{isinstance} function was the most common, observed in 26 cases, followed by the \texttt{lambda} mental set in 19 cases. The \texttt{enumerate} mental set appeared in 17 solutions, while the \texttt{sum} mental set was identified in 14.

\medskip
\subsection{Overcoming the Einstellung Effect (RQ2, H2)} 

The primary experimental phase of the study was completed by 39 participants. In the second set of tasks, two participants from the Control group demonstrated a shift from indirect to direct problem-solving strategies. As a result, we cannot definitively conclude that the intervention was effective in overcoming the mental set. The potential factors that may have contributed to these results will be discussed at length in \Cref{sec:discussion}.

\medskip
\subsection{Response Time Difference (RQ3, H3)} 

\begin{figure}[h]
    \centering
    \includegraphics[width=0.75\textwidth]{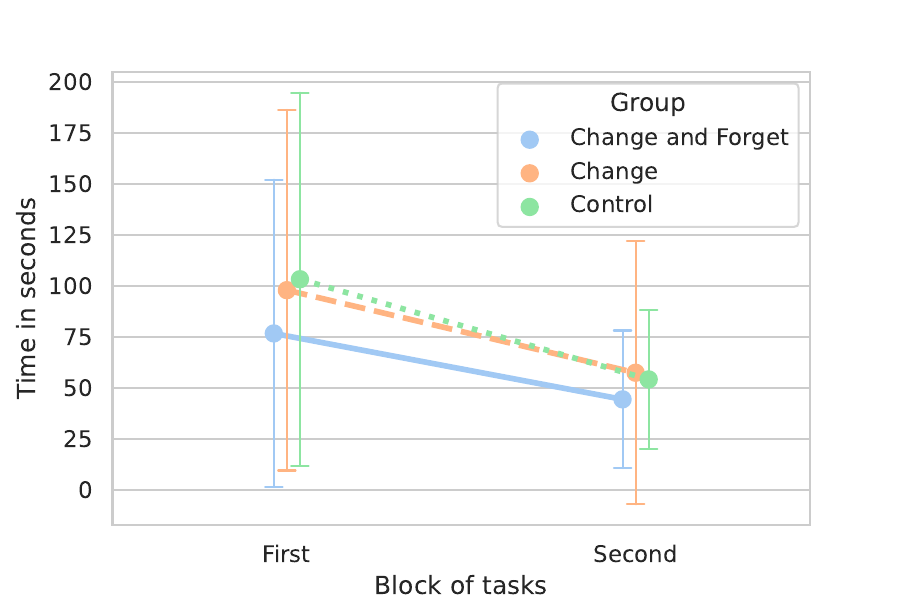}
    \caption{Mean response time for each set of tasks by the group of participants.}
    \label{fig:time}
\end{figure}

The mean response time plot in \Cref{fig:time} sheds light on the cognitive load experienced by participants during the experiment. Specifically, we examined the response time difference between corresponding tasks, which provided insights into the process of solution de-automatization in each group. Interestingly, our results indicate that the participants in the intervention groups (Change, Change and Forget) did not experience a significantly larger response time difference than those in the Control group (\APAh{2}{1.403}{.496}). Given that response time is widely used as a measure of cognitive load in experimental research~\shortcite{huang2009measuring, zheng2012}, our findings suggest that the intervention did not noticeably affect the de-automatization of habitual problem-solving strategies among the participants in the intervention groups.

\medskip
\subsection{Helpfulness and Comfortability of Intervention (RQ4, H4)} 

\begin{figure}
    \centering
    \includegraphics[width=0.5\textwidth]{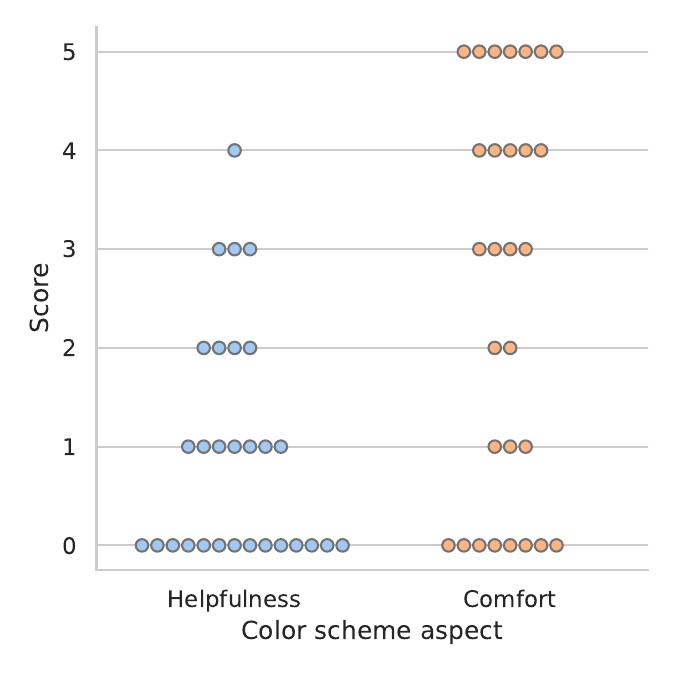}
    \caption{Comfortability and helpfulness scores of the color scheme change.}
    \label{fig:how}
\end{figure}

The plot shown in \Cref{fig:how} displays the participants' estimates of the helpfulness and comfortability of the intervention on a scale ranging from 0 to 5. Our analysis revealed that the participants perceived the color scheme change as unhelpful \APAstats{0.97}{1.18}. This is not surprising given that there was no significant change in the solutions. Additionally, we found that the comfortability level of the color scheme change varied among participants, as indicated by the relatively high standard deviation \APAstats{2.55}{2.01}. These results suggest that the interventions provided in our research may not be effective in overcoming the mental set and that in the future research it is important to consider individual differences which may play a role in how participants perceive the interventions.

With 29 participants in the intervention groups, we were unable to find a statistically significant correlation between the levels of helpfulness and comfortability (\APAr{29}{.3}{.36}). However, we did find a weak but significant positive correlation between the helpfulness scores and the difference in response time (\APAr{116}{.272}{.003}). On the other hand, we found no significant correlation between the comfortability and the difference in response time (\APAr{116}{.003}{.972}). These results suggest that there is a logical connection between how participants rated the helpfulness of the intervention and the observed change in solution speed after the treatment.

\medskip
\subsection{Insights from the Open-ended Question} 

In the final step of our study, we asked participants to explain their decision to change or maintain their problem-solving approach when presented with the problem the second time. Of the 39 participants, 22 made some changes to their solution, but the majority of these changes did not involve attempting to overcome the mental set. Rather, they were minor edits such as fixing typos or making minor code adjustments. Only 2 participants from the Control group demonstrated the successful mental set overcoming by recalling a more efficient function. Based on our qualitative analysis of the data, we conclude that the presented tasks may not have been challenging enough for some participants and that the mental set on such material could potentially be overcome without any intervention.

Here are some quotes from the gathered answers. One of the participants wrote: 
\observation{I went through the tasks quickly and didn’t even notice that tasks were repeated. It was less comfortable to think on a white background, it was more difficult to come up with a solution that I would like. Indeed, in some tasks, ideas of other solutions came, but in such simple tasks it was difficult to use a different solution. After all, there is a call to 1-2 functions of the standard library.} 

We also got such feedback: 
\observation{In one of the tasks of the second set, I decided that instead of taking up space with a ``for'' loop I can use a list expression to iterate through the elements, as well as using ``.join''. On the other hand, such code would be more difficult to understand when looking for errors.}

%% file: sections/05-discussion.tex
\section{DISCUSSION} \label{sec:discussion}

The primary objective of our study was to investigate the efficacy of two interventions, namely task-irrelevant change and intentional forgetting techniques, in aiding programmers in overcoming the  mental set (\eins) effect. 

Our study found that 44\% of the participants exhibited the mental set with the prevalence among less experienced programmers, as those who exhibited the mental set had significantly less experience with Python than those who did not. This finding aligns with the research on the relationship between expertise and problem-solving, which has shown that less experienced individuals often rely more heavily on habituated knowledge and less on analytical reasoning than experts~\shortcite{bilalic2008inflexibility,chi2006laboratory}. 

In terms of the types of mental set displayed, the \texttt{isinstance} mental set was the most common, followed by the \texttt{lambda} mental set. This finding suggests that these mental sets may be particularly ingrained in programmers' minds and could be difficult to overcome without explicit interventions. Future research could examine the effectiveness of different interventions in addressing these specific mental sets.

However, interventions provided in our research --- a task-irrelevant change in the developing environment and an instruction to forget the previously provided information --- were not efficient in terms of mental set overcoming. The main finding of our study is the fact that there was no switch between indirect and direct solutions in our experimental groups. Only two participants in the Control group exhibited a shift from indirect to direct problem-solving strategies in the second set of tasks. 

Several factors may have contributed to these results. 

It is probable that, despite the familiarity with both direct and indirect solutions, which was tested by our survey, the advantages of the direct solution were not explicit for all participants.

Previous studies on the impact of task-irrelevant change on the mental set have argued that the change disrupts the fixation by signaling the task as new, thereby prompting the generation of a fresh solution strategy~\shortcite{tuchtieva2011,tukhtieva2014influence}. However, these findings were obtained in studies involving numerical and mathematical tasks. In our research, the tasks were presented in a natural language and required writing several lines of code in a specific programming language. It is possible that the difference in the form and the level of task abstraction, specifically the lack of items to manipulate, contributed to the absence of the \eins effect overcoming.

It is also possible that the intervention was not strong enough to facilitate significant changes in problem-solving strategies. In previous research, the change in task-irrelevant parameters was much more noticeable than in our study. The change involved more visual noise and a different modality during presentation~\shortcite{tuchtieva2016,luchins1950new}. However, we intentionally avoided such a dramatic change to keep experimental conditions as close to real-life situations as possible. We do acknowledge that a change to a less common color scheme, as seen in Tukhtieva's work with bright and colorful photographs, could produce a more noticeable effect. However, implementing such a change in a code editor could harm its usability, as suggested by the low comfortability ratings in our study. One feasible solution to introduce a significant yet comfortable task-irrelevant change is to request users who are habituated to the solution, for example, in a Web-based Massive Open Online Courses, to switch from a Web-based coding environment to a standalone IDE.

Additionally, some studies have shown that the stability of a mental set can be attributed to the differences between the mental set formation environment and the intervention environment~\shortcite{chen2004schema}. As we did not include the step of mental set formation in our design, it is possible that due to that, the provided interventions did not affect the \eins effect.

Based on our results, it seems that the interventions did not have a substantial effect on the participants' cognitive load and conscious control, as evidenced by the response time difference between the corresponding tasks. Previous research has suggested that an increase in the cognitive load can be beneficial for conscious control, as it can help individuals pay more attention to the task at hand and solve it more efficiently, leading to better de-automatization and solution efficacy~\shortcite{allakhverdov2019cognitive,moroshkina2008soznatelnyy}. However, the results of our study indicate that this may not have been the case in our research. 

The results of our study suggest that the interventions we provided may not have been effective in breaking the participants' habitual problem-solving strategies. These findings emphasize the importance of further research aimed at exploring alternative approaches to promote de-automatization and facilitate creative problem-solving in the field of programming. 

%% file: sections/06-ttv.tex
\section{THREATS TO VALIDITY} \label{sec:ttv}

To ensure the validity of our findings, we considered several threats to validity.

\medskip
\subsection{Construct Validity}

The construct validity of this study refers to the extent to which our measurements accurately capture and represent the presence of the mental set and the effectiveness of the interventions employed. In other words, it assesses how well our chosen measurements align with the underlying concepts they are intended to measure, ensuring the reliability and accuracy of our findings.

The primary threat to construct validity of this study is the absence of a standardized measurement tool specifically designed for assessing the mental set phenomenon. Although we used a screening questionnaire and a programming task to identify participants with relevant mental sets, there is a possibility that our methods did not capture some participants with a mental set. However, it is important to note that we focused our attention on the four common mental sets in programming, which mitigates some concerns regarding our ability to capture all types of mental sets.

Another threat to the construct validity is the measurement of the interventions' effectiveness in overcoming the mental set. To address this, we used two complementary measures as a proxy for problem-solving de-automation: (a) the number of switches from indirect to direct solutions and (b) the response time difference between the sets of tasks.

It is also important to note that our sample size was relatively small and there is a possibility of the selection bias, as participants with a mental set may have been more likely to participate in the study to improve their problem-solving skills. This may limit the generalizability of our findings.

\medskip
\subsection{Internal Validity}

Internal validity refers to the extent to which our experimental design accurately identifies the effect of the intervention on the mental set. 

The primary threat to the internal validity of our study is the presence of alternative explanations for the observed results. We used a randomized controlled design to minimize the impact of confounding variables. To mitigate the risk that the improvement of participants' performance was due to task repetition, rather than the interventions, we used two different problems with different structures, rather than repeating the same task.  Additionally, we carefully arranged and varied the sequence in which participants received the different interventions to control for order effects. Also, we attempted to control for potentially confounding variables such as age and education level, by selecting a diverse sample. However, it is possible that there were other unmeasured variables that could have impacted our results.

\medskip
\subsection{External Validity}

External validity refers to the extent to which our findings can be generalized to other contexts. Our study was conducted on a sample of Python programmers who voluntarily participated in the study. Therefore, our results may not generalize to programmers who use different programming languages. Moreover, our sample included participants who are more likely to be motivated to learn and improve their programming skills, which may not be representative of all programmers. To increase the generalizability of our results, future research should replicate our study with participants from different programming communities or with different problem structures.

%% file: sections/07-conclusion.tex
\section{CONCLUSION} \label{sec:conclusion}

In this work, we explore the intersection of cognitive psychology and software engineering. Specifically, we address the issue of cognitive rigidness in programming. 

We provide evidence that the mental set (\eins) effect, which refers to the tendency to stick to habitual solutions rather than exploring new ones during problem-solving, is prevalent among Python developers who have not exceeded their second year of familiarity with the language. Our study examines the effectiveness of two interventions --- task-irrelevant change and intentional forgetting technique --- in overcoming the \eins effect. Although participants reported engagement in the rethinking process, our findings suggest that these interventions were insufficient in facilitating a switch from indirect to direct solutions. We suggest that a more dramatic and unhabitual change of the working environment, such as altering font size/type or switching from a web-based to a standalone IDE, may be helpful in overcoming cognitive rigidity.

Our study contributes to the existing literature on creativity support during problem-solving in software development and provides a framework for future experimental research. By sharing our open-source code and data, we offer resources for further investigations in this field. We believe that studies combining cognitive psychology and programming, like this one, can shed light on ways to help practitioners to be more efficient in the field of software engineering.